\documentclass[a4paper,pra,aps,twocolumn,10pt,superscriptaddress]{revtex4-1}

\usepackage{amssymb,amsmath}
\usepackage{graphicx}
\usepackage{hyperref}

\begin{document}

\title{The effect of a laser dip in the semiclassical dynamics of bosonic Josephson Junctions}
\author{G. Szirmai}
\affiliation{Institute for Solid State Physics and Optics - Wigner Research Centre for Physics, Hungarian Academy of Sciences, H-1525 Budapest P.O. Box 49, Hungary}
\author{G. Mazzarella}
\affiliation{Dipartimento di Fisica e Astronomia ``Galileo Galilei'' and CNISM, Universit\`a di Padova, Via Marzolo 8, I-35131 Padova, Italy}
\author{L. Salasnich}
\affiliation{Dipartimento di Fisica e Astronomia ``Galileo Galilei'' and CNISM, Universit\`a di Padova, Via Marzolo 8, I-35131 Padova, Italy}

\begin{abstract}
We consider the standard double well setup extended with a laser beam in the center to create a ``triple well" potential. The beam in the center is much more narrow than the barrier, and it creates a tunable depth well which can support a localized state in the middle. We show that the presence of the localized state in the central well changes the sign of tunneling between the left and right wells and therefore controls the fixed point dynamics of the bosonic Josephson junction.
\end{abstract}

\maketitle

\section{Introduction}

A Bose-Einstein condensate of a dilute gas of alkaline atoms in a double well potential realizes the physics of Josephson junctions, which was originally predicted in two superconductors separated by an insulating layer \cite{josephson1962possible}. The bosonic realization of Josephson junction physics has attracted a great interest both theoretically \cite{javanainen1986oscillatory, smerzi1997quantum, raghavan1999coherent, milburn1997quantum, mazzarella2010spontaneous, mazzarella2011coherence, julia2010macroscopic, julia2010bose, gillet2014tunneling} and experimentally \cite{albiez2005direct, levy2007ac, zibold2010classical} in recent years. On one hand the physics of Josephson junctions can be described by the two coupled nonlinear equations of a non-rigid pendulum, therefore its careful investigation is very tempting, since the model and its mathematics look fairly simple, while they are complicated enough in order to help us understanding some aspects of more elaborate problems, like the Bose-Hubbard model. In particular, bosonic Josephson junctions (BJJs) may be regarded as a two-site realization of the Bose-Hubbard model. On the other hand the mesoscopic coherent dynamics of Bose-Einstein condensate has important issues of its own, such as the validity of semiclassical dynamics and the use of coherent states in few mode and finite atom number systems \cite{mazzarella2011coherence, julia2010macroscopic}.

The tunneling dynamics of BJJs can serve as a basic tool in interferometry applications \cite{shin2004atom, shin2005interference, schumm2005matter}. The first experiments with repulsively interacting Bose condensates revealed self-trapping and plasma oscillations \cite{albiez2005direct} and later, with an experimental effort the a.c. Josephson effect was also observed \cite{levy2007ac}. With the help of atomic Feshbach resonances it is possible to change the magnitude and even the sign of the parameter of on-site interaction. Therefore it is in principle possible to ``quench'' the dynamics of the BJJ and realize the semiclassical dynamics around the stationary points of the Josephson equations or change the dynamics governed by one particular fixed point to a different one governed by a different fixed point \cite{zibold2010classical}. This way a setup for very fast macroscopic entanglement generation can be achieved \cite{micheli2003many,Vidal2004Entanglement}.

The question naturally arises whether it is possible or not to obtain some similar quenching not only with the on-site interaction, but rather by engineering the tunneling amplitude of the junction? In this paper we give an affirmative answer to this question. With the help of an external, tightly focused, red-detuned laser beam one can create a tiny hole in the middle of the double-well barrier. When the depth of this dip is increased, at some point a bound state localized inside the dip potential appears, and by further increasing the potential depth the tunneling constant between the original left and right wells changes sign.  The creation of such a static obstacle is fairly simple and therefore gives another knob on the system besides the standard Feshbach resonance technique.

The plan of the paper is as follows. In Sec. \ref{sec:dwd} we consider the single particle problem where a dip potential is superimposed on the standard double well. In Sec. \ref{sec:bjj} we apply the two-mode approximation to the problem when the doublet formed by the Wannier states of the left and right wells are sufficiently separated from the other energy levels and consider the Josephson dynamics. We summarize in Sec. \ref{sec:sum}. The stability analysis of the stationary points of the dynamics is moved to the Appendix.

\section{Double well with a dip in the middle}
\label{sec:dwd}

\begin{figure*}[ht!]
\centering
\includegraphics[width=\textwidth]{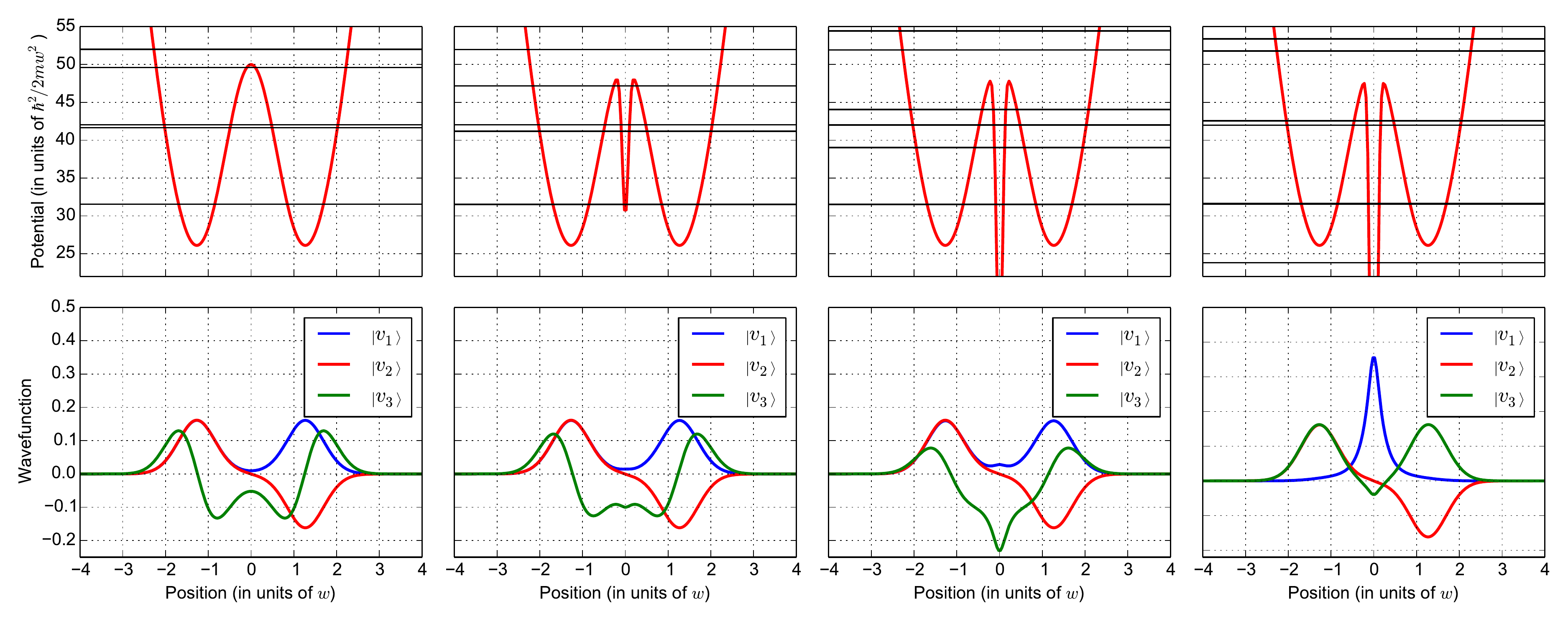}
\caption{(Color online) Top panel: the potential landscape with the energy eigenvalues for various laser intensities $I_0$. Bottom panel: the wave functions corresponding to the lowest three energies. $v_1(x)$ is the ground state wave function, $v_2(x)$ is the wave function of the first excited state, and $v_3(x)$ is the second excited state. From left to right the parameter $I_0$ varies as: $0.0, 2.0, 4.0, 8.0$}
\label{fig:lowest_states}
\end{figure*}
The double well setup considered here consists of a symmetric potential
\begin{equation}
\label{eq:dwpot}
V_{\text{DW}}(x)=\frac{1}{2}m\omega_H^2 x^2+V_1\, e^{-\frac{x^2}{2 w^2}},
\end{equation}
where $m$ is the mass of the atoms, $\omega_H$ is the frequency of the parabolic confinement, $V_1$ is the height and $w$ is the width of the double well barrier. We consider tight confinement in the perpendicular directions and treat the system as one-dimensional. In addition to the double well potential there is a tightly focused laser beam in the center which is red detuned from the atomic transition creating a further attractive potential for the atoms,
\begin{equation}
\label{eq:centerwell}
V_{L}(x)=-I_0\,e^{-\frac{x^2}{2 \sigma^2}}
\end{equation} 
where $I_0$ is the strength and $\sigma\ll w$ is the width of the optical potential. The full single particle Hamiltonian is
\begin{equation}
\label{eq:ham}
\hat H=-\frac{\hbar^2}{2m}\frac{d^2}{dx^2} + V_{\text{DW}}(x)+V_L(x).
\end{equation}
The perturbing potential $V_L(x)$ opens up a  narrow dip in the center of the double well barrier, as illustrated in Fig. \ref{fig:lowest_states}. When varying the strength $I_0$, one can interpolate between a symmetric double well potential and a triple well one. For $I_0=0$, with our choice of parameters (for ${}^{87}\mathrm{Rb}$), which is close to experimental applications ($m=87\,\mathrm{amu}$, $\omega_H=2\pi\times15\,\mathrm{Hz}$,  $w=5\,\mathrm{\mu m}$,  $V_1=5 m \omega_H^2 w^2$, and $\sigma=0.5\,\mathrm{\mu m}$) the lowest two energy eigenvalues are almost degenerate and they form the low energy doublet of the double well problem. The corresponding wave functions are the symmetric and antisymmetric combinations of the Wannier orbits, which themselves are localized states around the left and right energy minima of the potential. Other energy eigenvalues are much higher and one can rely on a two-mode approximation when treating the problem.
\begin{figure}[b!]
\centering
\includegraphics[width=\columnwidth]{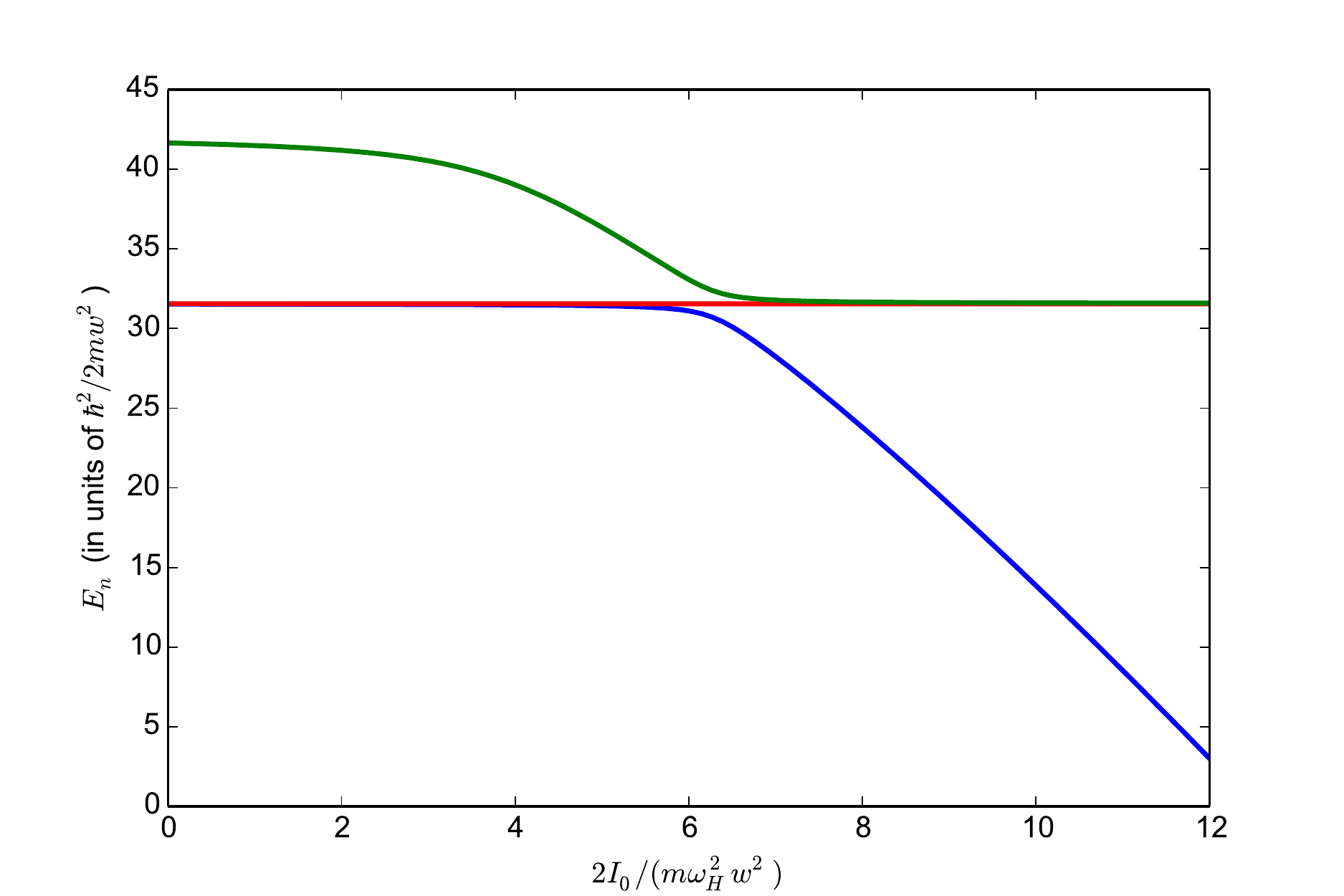}
\caption{(Color online) The three lowest energy eigenvalues plotted as a function of $I_0$. One can observe an avoided crossing. The second energy level is unaffected by the perturbing potential, while the lowest energy and the third energy eigenvalue tilt down with increasing $I_0$.}
\label{fig:energy_crossing}
\end{figure}

When $I_0$ is increased gradually, as shown in the subsequent plots in Fig \ref{fig:lowest_states},  a central well starts to form in the middle of the potential barrier. For small values of $I_0$ the central well doesn't support a localized state and its effect is just a small perturbation of the energy eigenvalues and an even smaller one on the wave functions. The three lowest energy eigenvalues are plotted in Fig. \ref{fig:energy_crossing}. One eigenvalue of the doublet is basically unchanged by the perturbation, namely the one which corresponds to the antisymmetric wave function, which has a node at the position of the perturbation. The other eigenvalue is shifted a little bit downwards. As $I_0$ increases, the central well deepens, and the third energy eigenvalue approaches the low energy doublet. As this third energy eigenvalue comes closer and closer, the two-mode description becomes more and more inaccurate. One can observe an avoided crossing in the three lowest energy eigenvalues. For small values of $I_0$ the lowest two eigenvalues form the doublet of the symmetric and antisymmetric combinations of the Wannier orbits. On the other side of the crossing, i.e. for large values of $I_0$, the single lowest energy eigenvalue correspond to the state localized in the central well, while the next two eigenvalues form now the doublet of the antisymmetric and symmetric combinations of the Wannier orbits localized at the left and right valleys. 

\section{Bosonic Josephson Junction}
\label{sec:bjj}

When the splitting of the low energy doublet is much smaller than the energy difference between the doublet and the closest other energy eigenvalue, the two-mode approximation gives a sufficiently accurate description of the tunneling dynamics between the left and right wells. In this limit the other states are non-resonant and energy conservation decouples them from the tunneling dynamics. With the present parameters it means approximately either $I_0<I_{c,1}\approx6$, or $I_0>I_{c,2}\approx7$.

\begin{figure}[b!]
\centering
\includegraphics[width=\columnwidth]{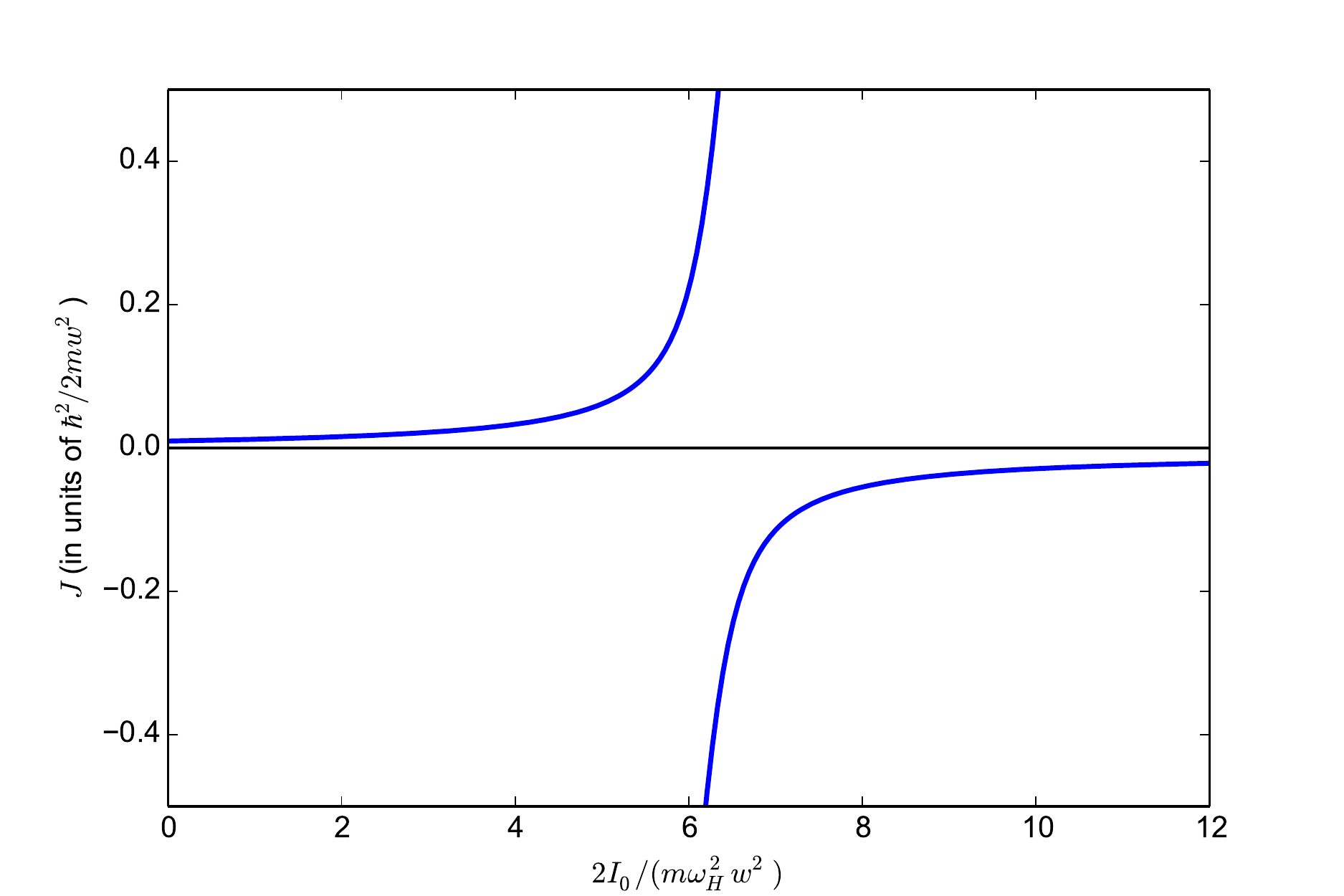}
\caption{(Color online) The tunneling ratio as a function of the depth of the central well, $I_0$.}
\label{fig:Jres}
\end{figure}

When $I_0<I_{c,1}$ the Wannier functions are given by: $w_1(x)=(v_1(x)+v_2(x))/\sqrt{2}$,  $w_2(x)=(v_1(x)-v_2(x))/\sqrt{2}$, for the left and right wells, respectively. For $I_0>I_{c,2}$ the first and second excited states give the Wannier functions, and they read as:  $w_1(x)=(v_2(x)+v_3(x))/\sqrt{2}$,  and $w_2(x)=(v_2(x)-v_3(x))/\sqrt{2}$, for the left and right wells, respectively. In second quantized form the non-interacting Hamiltonian \eqref{eq:ham} can be cast to the following form:
\begin{equation}
\label{eq:hamfree}
\hat H_0=\epsilon\left(\hat b_1^\dagger \hat b_1+\hat b_2^\dagger \hat b_2\right)-J\left(\hat b_1^\dagger \hat b_2+\hat b_2^\dagger \hat b_1\right),
\end{equation}
where the parameters are given by $\epsilon=\left< w_1\right|\hat H\left| w_1\right>$, and $J=-\left< w_1\right|\hat H\left| w_2\right>$. In the two-mode approximation the total atom number $\hat N=\hat b_1^\dagger \hat b_1+\hat b_2^\dagger \hat b_2$ is a constant of motion, therefore the first term in Eq. \eqref{eq:hamfree} can be dropped. The parameter $J$ shows a ``resonance" like behavior as a function of $I_0$, as illustrated in Fig. \ref{fig:Jres}. We note that the central part of the figure, where the crossing of the energy levels takes place, is not reliable, since the two-mode approximation breaks down. Nevertheless, the tunneling amplitude changes sign at the crossing and the lower energy orbital of the doublet changes from ungerade to gerade symmetry.

\begin{figure*}
\centering
\includegraphics[width=\textwidth]{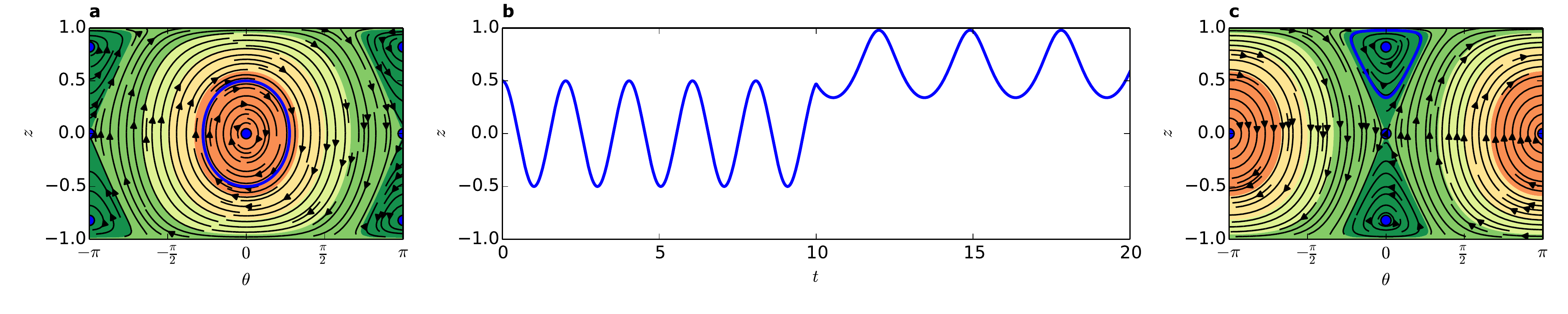}
\caption{(Color online) Time evolution of the solution of the BJJ equations when a laser dip is abruptly turned on at $t=10$ (time is measured in units of $|J|^{-1}$). For $t<10$ the dip potential is switched off and the parameters are $U N/J=3.5$. The initial conditions are $(z(0)=0.5, \theta(0)=0)$. The system performs Josephson oscillations. At $t=10$ a dip potential is suddenly turned on and kept constant. For $t>10$ the parameters change to $U N/J=-3.5$ and the dynamics exhibits self-trapping. b) shows the population imbalance as a function of time. Subfigure a) shows, for $t<10$,  the phase space, fixed points ($\mathbf{X}_1$ and $\mathbf{X}_4$), and the  oscillation (thick line) corresponding to the initial conditions. Panel c) shows the phase space for the new system parameters valid from $t=10$, the fixed points $(\mathbf{X}_2$ and $\mathbf{X}_3$). The thick line corresponds to the new trajectory of the system continuing its oscillation in the new energy landscape.}
\label{fig:quench}
\end{figure*}
In the presence of interaction the Hamiltonian is modified to $\hat H=\hat H_0 + \hat H_I$, with
\begin{equation}
\hat H_I=\frac{U}{2}\left(\hat b_1^\dagger\hat b_1^\dagger\hat b_1\hat b_1+\hat b_2^\dagger\hat b_2^\dagger\hat b_2\hat b_2\right),
\end{equation}
where $U$ characterizes the on-site interaction. At sufficiently low temperatures the bosons form a Bose-Einstein condensate, and in the semi-classical approximation the atomic operators are replaced with c-numbers: $b_k=\sqrt{N_k}(t)e^{i\theta_k(t)}$, where $N_k(t)$ is the atom number in well $k$ at time $t$, and $\theta_k(t)$ is the corresponding phase. The total atom number  $N_1(t)+N_2(t)\equiv N$ is constant. It is convenient to introduce the fractional population difference of the two wells, $z(t)=[N_1(t)-N_2(t)]/N$, and the relative phase $\theta(t)=\theta_2(t)-\theta_1(t)$. Using this substitution in the Hamiltonian $\hat H$ one can arrive to the semi-classical energy function \cite{smerzi1997quantum}
\begin{equation}
\label{eq:semen}
\mathcal{H}(z,\theta)=-2JN\sqrt{1-z^2}\cos(\theta)+\frac{U}{2}N^2z^2,
\end{equation}
from which the semi-classical equations, known as the bosonic Josephson junction equations can be derived as
\begin{subequations}
\label{eqs:BJJ}
\begin{align}
\dot z&=-\frac{1}{N}\frac{\partial \mathcal{H}}{\partial \theta}=-2J\sqrt{1-z^2}\sin(\theta),\\
\dot\theta&=\frac{1}{N}\frac{\partial \mathcal{H}}{\partial z}=\left(U\,N+\frac{2J}{\sqrt{1-z^2}}\cos(\theta)\right)z.
\end{align}
\end{subequations}
Here and from now on we work with $\hbar=1$. Equations \eqref{eqs:BJJ} have 4 stationary solutions $\dot{\bar{z}}=0$ and $\dot{\bar{\theta}}=0$: It has two zero imbalance solutions with $\mathbf{X}_1=(\bar z=0, \bar \theta=0)$, and  $\mathbf{X}_2=(\bar z=0, \bar \theta=\pi)$. Furthermore there are two finite imbalance solutions: $\mathbf{X}_3=(\bar z=\sqrt{1-(2J/U N)^2}, \bar \theta=0)$, and  $\mathbf{X}_4=(\bar z=\sqrt{1-(2J/U N)^2}, \bar \theta=\pi)$. By substituting the stationary solutions to the semi-classical energy function \eqref{eq:semen}, one can immediately see, that for $U>0$ the zero imbalance solutions always have the lowest energy. Also depending on the sign of $J$ the minimal energy solution is either with $\bar\theta=0$ for $J>0$, and $\bar\theta=\pi$ for $J<0$. For attractive interaction $U<0$ the finite imbalance solutions are energetically more favorable for $(UN)^2>4J^2$, and the tunneling dynamics exhibits self-trapping \cite{raghavan1999coherent}. Thus, points of $(\bar{z},0)$ with $\bar{z} \neq 0$ are stable fixed points for the ODEs \eqref{eqs:BJJ} only in the presence of attractive on-site interactions $U$ provided that $U<-2|J|/N$. Under inital conditions $(z(0),0)$  -  with $z(0)<(2/\Gamma) (\Gamma-1)^{0.5}$ ($\Gamma=|UN/2J|$) - the solutions of these ODEs describes oscillations of the fractional imbalance and relative phase about a nonzero time  averaged value and zero, respectively.

By suitably tuning $I_0$,  one can change the sign of $J$ by moving from the left side of the resonance to the right side of it (see Fig. \ref{fig:Jres}). All the above condition thus can be satisfied and one can quench between self-trapping  and Josephson dynamics (and vice versa),  even with repulsive boson-boson interaction. In Fig. \ref{fig:quench} we illustrate the quench dynamics for a repulsive Bose condensate prepared initially for $(z(0)=0.5, \theta(0)=0)$. At $t=0$ the the dip potential is turned off and we have a symmetric double well potential with $J>0$. The system starts Josephson (plasma) oscillations. In panel (a) we show the phase space trajectories and fixed points of the semi-classical Hamiltonian \eqref{eq:semen} for $U N=3.5 J$. The shading corresponds to the energy, where the central (orange) region is the energy minimum and the outer (green) regions correspond to higher energies. The thick line shows the trajectory of the initial Josephson oscillation. On panel (b) we show the population imbalance as a function of time, measured in units of $|J|^{-1}$. The system parameters are left unchanged for $t=10 J^{-1}$. At $t=10 J^{-1}$ we switch on abruptly a dip potential with $I_0$ such to go to the other side of the resonance with $J\rightarrow -J$. Now the phase space diagram is depicted in panel (c). As we see, due to the change of the sign of $J$, the energy landscape changes by $\theta\rightarrow \theta+\pi$, and the finite imbalance (unstable) fixed points corresponding to the energy maxima are moved to the center. The Bose condensate continues its dynamics in the modified landscape, around the $\mathbf{X}_3$ fixed point, which is selected by its instantaneous state $(z(10|J|^{-1}), \theta(10|J|^{-1}))$. This self-trapping dynamics is shown also in panel (b) for $t>10 |J|^{-1}$.

Another indicator of the change of the type of the dynamics is the change in the oscillation frequency, which (at least for small oscillations around the fixed points) can be calculated by the linear stability analysis of the fixed points, as summarized in Appendix \ref{sec:stabanal}. In Fig. \ref{fig:freq} we plot the oscillation frequency as a function of $I_0$.  As we increase $I_0$ at the left hand side of the resonance, the Josephson oscillation frequency $\omega_J$ starts to grow first, since $J$ increases, and then at the right hand side where $J<0$ it decreases again, since $|J|$ decreases. Then at some point, when $U$ becomes bigger than $2|J|$ the fixed point for the Josephson oscillation becomes unstable and instead the self-trapping frequency $\omega_{\text{ST}}$ appears.

\begin{figure}[b!]
\centering
\includegraphics[width=\columnwidth]{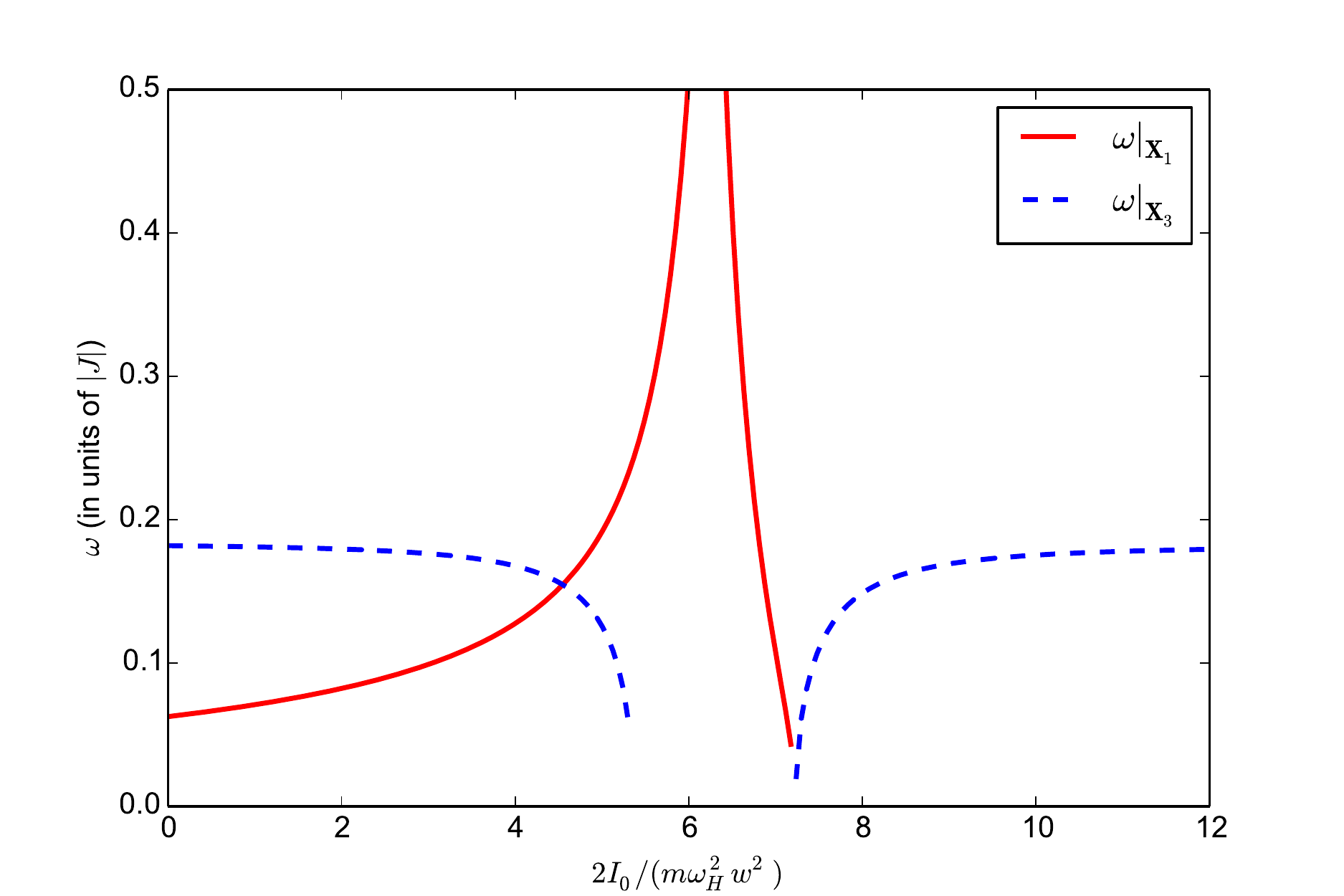}
\caption{(Color online) The oscillation frequencies as function of the strength of the potential dip. The given fixed point becomes unstable, when the curve goes below zero.}
\label{fig:freq}
\end{figure}

\section{Summary}
\label{sec:sum}

In this paper we have considered the effect of an additional central well added to the center of the symmetric double well barrier. We have shown that by suddenly opening up this narrow central well the tunneling amplitude of the bosonic Josephson junction can be ``quenched'' to almost arbitrary values. Therefore in experiments one can have an additional tunable parameter on the double well system and change the dynamics in-situ from plasma oscillations to the a.c Josephson dynamics or even to self-trapping without modifying the scattering properties.

\section*{Acknowledgements}

GSZ acknowledges support from the Hungarian National Office for Research and Technology under the contract ERC\_HU\_09 OPTOMECH, the Hungarian Academy of Sciences (Lend\"ulet Program, LP2011-016), the Hungarian Scientific Research Fund (grant no. PD104652) and the J\'anos Bolyai Scholarship. GM and LS acknowledge financial support from Università di Padova (Progetto di Ateneo grant No. CPDA 118083), Cariparo Foundation (Eccellenza grant 2011/2012), and MIUR (PRIN grant No. 2010LLKJBX).

\appendix
\section{Fixed point stability}
\label{sec:stabanal}

In order to check the stability of the solutions, we look for small perturbations around the stationary points and calculate the linear stability matrix of Eqs. \eqref{eqs:BJJ}
\begin{equation}
\label{eq:stabmat}
\left(\begin{array}{c}
\delta \dot z\\
\delta \dot \theta
\end{array}\right)=\left(
\begin{array}{c c}
\frac{2J\bar z\sin(\bar\theta)}{\sqrt{1-\bar z^2}}&-2J\sqrt{1-\bar z^2}\cos(\bar\theta)\\
UN+\frac{2J\cos(\bar\theta)}{(1-\bar z^2)^{3/2}}&-\frac{2J\bar z\sin(\bar\theta)}{\sqrt{1-\bar z^2}}
\end{array}\right)
\left(\begin{array}{c}
\delta z\\
\delta \theta
\end{array}\right).
\end{equation}
The linear stability matrix has the following eigenvalues:
\begin{equation}
\label{eq:evals}
\lambda=\pm i\sqrt{4 J^2\left[\frac{\cos(2\bar\theta)}{1-\bar z^2}+\frac{UN}{2J}\sqrt{1-\bar z^2}\cos(\bar\theta)\right]}.
\end{equation}
For purely imaginary eigenvalues, the stationary solution is marginally stable: small perturbations around the solution result in periodic oscillations. The frequency of the oscillation is, $\omega=\mathrm{Im}\,\lambda$. On the other hand, when the quantity under the square root becomes negative, the eigenvalues become a pair of real numbers with equal magnitude and opposite sign, and the perturbations can exponentially grow in time. By directly substituting the stationary solutions to the eigenvalues we get
\begin{subequations}
\label{eqs:evalevals}
\begin{align}
\lambda|_{\mathbf{X}_1}&=\pm i \sqrt{4J^2\left(1+\frac{U N}{2 J}\right)},&\text{stable if: } &\frac{U N}{2 J}>-1,\\
\lambda|_{\mathbf{X}_2}&=\pm i \sqrt{4J^2\left(1-\frac{U N}{2 J}\right)},&\text{stable if: } &\frac{U N}{2 J}<1,\\
\lambda|_{\mathbf{X}_3}&=\pm i \sqrt{(U N)^2-4J^2},&\text{stable if: } &(U N)^2>4J^2,\\
\lambda|_{\mathbf{X}_4}&=\pm i \sqrt{(U N)^2-4J^2},&\text{stable if: } &(U N)^2>4J^2.
\end{align}
\end{subequations}
For the zero imbalance solutions, $\mathbf{X}_1$ and $\mathbf{X}_2$, the frequency $\omega$ is the Josephson
frequency $\omega_\text{J}$. During the dynamics around $\bar{z} = 0$ there is population inversion, i.e. $z(t)$ changes sign. Instead, for $\mathbf{X}_3$ and $\mathbf{X}_4$,  when $\bar{z} \neq 0$, this frequency is the self-trapping frequency $\omega_{\text{ST}}$; during the dynamics there is no population inversion, i.e. $z(t)$ does not change sign.

\bibliography{dwlaser}

\end{document}